\documentclass[aps,pre,twocolumn,showpacs]{revtex4}
\usepackage{epsfig}
\usepackage{times}
\begin{document}

\title{Dynamically generated cyclic dominance in spatial prisoner's dilemma games}

\author{Attila Szolnoki$^1$, Zhen Wang$^{2,3}$, Jinlong Wang$^2$, and Xiaodan Zhu$^2$}
\affiliation
{$^1$Research Institute for Technical Physics and Materials Science,
P.O. Box 49, H-1525 Budapest, Hungary \\
$^2$School of Innovation Experiment, Dalian University of Technology, Dalian 116024, China\\
$^3$School of Mathematical Science and College of Advanced Science and Technology, Dalian
University of Technology, Dalian 116024, China}

\begin{abstract}
We have studied the impact of time-dependent learning capacities of players in the framework of spatial prisoner's dilemma game. In our model, this capacity of players may decrease or increase in time after strategy adoption according to a step-like function. We investigated both possibilities separately and observed significantly different mechanisms that form the stationary pattern of the system. The time decreasing learning activity helps cooperator domains to recover the possible intrude of defectors hence supports cooperation. In the other case the temporary restrained learning activity generates a cyclic dominance between defector and cooperator strategies, which helps to maintain the diversity of strategies via propagating waves. The results are robust and remain valid by changing payoff values, interaction graphs or functions characterizing time-dependence of learning activity. Our observations suggest that dynamically generated mechanisms may offer alternative ways to keep cooperators alive even at very larger temptation to defect.
\end{abstract}

\maketitle

\section{Introduction}

The possible survival of cooperator strategy when defection pays more is a widely studied puzzle that is in the interest of very different disciplines from biology and physics \cite{szabo_pr07} to economy and social sciences \cite{hofbauer_98}. In the framework of evolutionary game theory several new mechanisms and player specific conditions were identified in the last decade that promote cooperation among selfish individuals \cite{nowak_s06,traulsen_jtb07,helbing_pre10}. Without listing all of them, we should mention the seminal observation of Santos and Pacheco \cite{santos_prl05} who realized that the heterogeneity of the interaction graph can amplify the mechanism of so-called network reciprocity \cite{nowak_n92b}. To follow this avenue, other types of heterogeneity of players were also identified which also increase the chance for cooperators to keep alive. These are the heterogeneity of strategy teaching activity \cite{szolnoki_epl07}, social diversity \cite{perc_pre08b,santos_n08}, heterogeneous influential effect \cite{wu_zx_cpl06}, variousness of payoffs \cite{perc_njp06c} or payoff allocation \cite{peng_d_epjb10}. Generally, the common mechanism behind these conditions is the enhanced possibility to form homogeneous cooperator clusters against the invasion of defectors.

Heterogeneity itself, however, is not a sufficient condition for promoting cooperation if the above mentioned mechanism cannot work. To give an example, the heterogeneity of strategy learning capacity does not support cooperation notably \cite{szolnoki_epl07}. Accordingly, we do not expect relevant change by allowing different strategy learning skills for players during strategy imitation process. A conceptually different situation emerges when a quantity that determines imitation of neighbors is time-dependent during the evolution of strategies. This extended freedom of the evolutionary process, as a kind of coevolutionary model \cite{perc_bs10}, may result in spontaneous evolution to a cooperator promoting heterogeneous state \cite{szolnoki_njp08}, but may also offer dynamically maintained mechanisms to help cooperators. For example, the age related teaching activity of players, when player's teaching activity increases with elapsed time after strategy pass, resulted a new type of dynamical mechanism that based on biased propagation of different strategies \cite{szolnoki_pre09}. More precisely, a highly selective promotion of cooperator-cooperator and defector-defector pairs emerges, which hinders influential defectors from spreading their strategy effectively across the interaction graph. On the other hand, a similar blocking mechanism does not exist around aged cooperators, who can spread cooperation via supporting cooperator newborns in the neighborhood.
As we argued earlier, similar mechanism cannot be expected for time-dependent strategy learning capacity of players because this type of diversity is unable to amplify network reciprocity.

The changing learning capability of players, however, is a viable assumption that should be considered. As we will argue in the next section, both the time-decreasing and time-increasing learning capacity of players are reasonable in evolutionary game models. By using the payoff elements of prisoner's dilemma, as the most challenging social dilemma, it turned out that the mentioned time dependences of learning skill have very different impacts on the stationary state. While time-decreasing learning capacity has a modest, but unambiguous cooperator promoting impact, the time-increasing learning skill results in a new type of state that can only be previously observed if more than two strategies were present in the system \cite{hauert_s02}. Although the change between the model definitions is tiny but the emerging mechanisms that form the stationary patterns are completely different. In the first case a sort of recovery mechanism helps cooperator domains to protect against the invasion of defectors, while in the second model a dynamically generated cyclic dominance maintains the diversity of strategies, even at hard conditions when cooperators cannot survive normally.

The remainder of this paper is organized as follows. In the next section we describe the studied models by also briefly discussing some possible extensions. In Sec.~\ref{A} we present the results of the first mentioned model. The result of the other type of time-dependence for learning capacity is presented in Sec.~\ref{B}. Finally, we summarize our observations and discuss their potential implications.

\section{Models}
\label{def}

For the sake of simplicity we study the weak version of prisoner's dilemma game that is characterized with the single variable parameter of temptation to defect $T=b$, the payoff received by a defector if playing against a cooperator, while the remaining payoff values are fixed. These elements are the reward for mutual cooperation $R=1$, the punishment for mutual defection $P=0$, and the suckers payoff $S=0$ received by a cooperator if playing against a defector \cite{nowak_n92b}. We should note here, however, that our main observations remain intact if real prisoner's dilemma is considered by using $S<0$ values. On the other hand, the temptation value is extended to $0 < b < 2$ interval to explore the stag-hunt region, too.

Keeping the model as simple as possible, the interaction graph of players is supposed to be a $L \times L$ square lattice where every player interacts with its four nearest neighbors by using the previously defined payoff elements. (The extension to other types of interaction graphs is also discussed in Sec.~\ref{B}.) Initially, each player $x$ on the graph is designated as a cooperator ($s_x = C$) or defector $D$ with equal probability. During the evolutionary process a randomly selected player $x$ acquires its payoff $p_x$ by playing the game with its nearest neighbors separately. Next, one randomly chosen neighbor, denoted by $y$, also acquires its payoff $p_y$ by playing the game with its four neighbors. Finally, player $x$ tries to enforce its strategy $s_x$ on player $y$ in accordance with the probability
\begin{equation}
W(s_x \rightarrow s_y)=w_y \frac{1}{1+\exp[(p_y-p_x)/K]},
\label{eq:prob}
\end{equation}
where $K$ denotes the amplitude of noise \cite{szabo_pre98}  
and $0 < w_y \le 1$ characterizes the strategy learning capability of player $y$. 

The latter parameter characterizes the willingness of player $y$ how she/he easily adopts the strategy of a neighbor independently on the payoff differences. This feature of players should not be necessarily constant in time but may change. Accordingly, two basically different situations can be assumed. In the first case (called model A) the learning capability of a player decreases as a function of $h_y$ elapsed time after a strategy adoption of $y$ player. (The quantity $h_y$ can be considered as an ``effective age'' of a strategy for player $y$.)
This time dependence describes the situation when a player who keeps a strategy unchanged for a long time may loose her/his capability to change. In other words, there is a constraint not to change a strategy that proved to be successful for long, despite the fact that a neighbor may be more successful presently. In the second case (called model B) the learning capacity of a player is an increasing function of elapsed time after strategy adoption. This time profile can be interpreted in a way that a player who just changed strategy is satisfied at first and is reluctant to change for a certain period. Later his/her willingness to adopt strategy is recovered again.

Naturally, there are several possibilities to link the personal $w_y$ learning capacity with the value of $h_y$, that is the elapsed time after a strategy adoption of player $y$. Without loosing generality, we will use simple step-like functions for both models. More precisely, for model A the learning capacity is given by
\begin{equation}
\label{modelA}
w_y = \cases {w_{max}=1,\,\,\,\,\, $if$ \,\,\,h_y \le \,\,$H$ \cr
              w_{min}, \,\,\,\,\,\,\,\,\,\,\,\,\,\,\,\,\, $otherwise$ \cr }\; ,
\end{equation}
where H is a sort of threshold value of iteration time when the learning capacity of focal player changes significantly. For a minimal learning capacity we choose $w_{min}=0.01$ to avoid the system being trapped in an intermediate frozen state. Evidently, for $w_{min}=1$ or for H~=~0 the original spatial model is restored where all players have uniform, time-independent learning capacities. The behavior of the latter model has already been studied extensively \cite{szabo_pre05}. 

\begin{figure}
\centerline{\epsfig{file=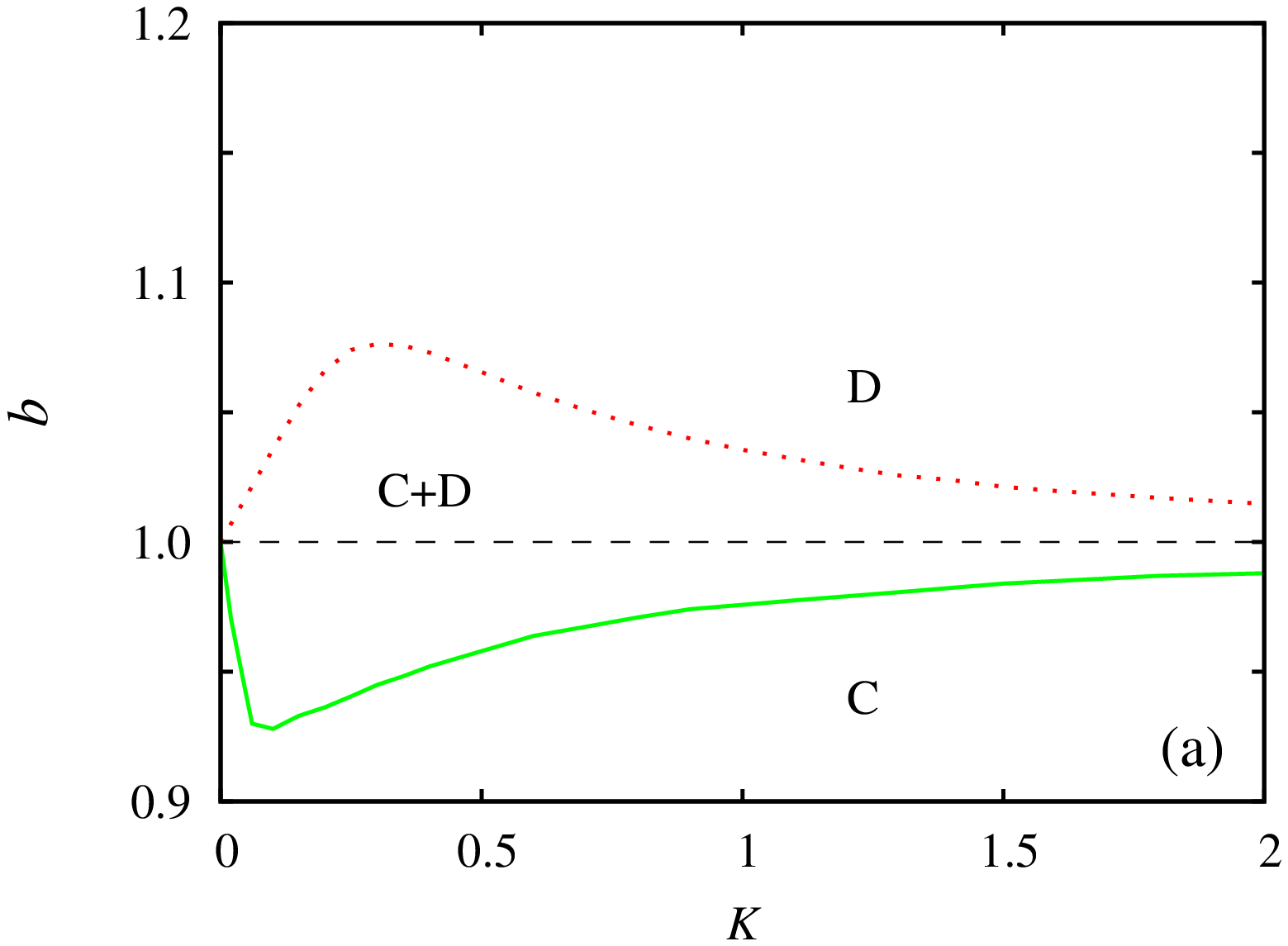,width=7.5cm}}
\centerline{\epsfig{file=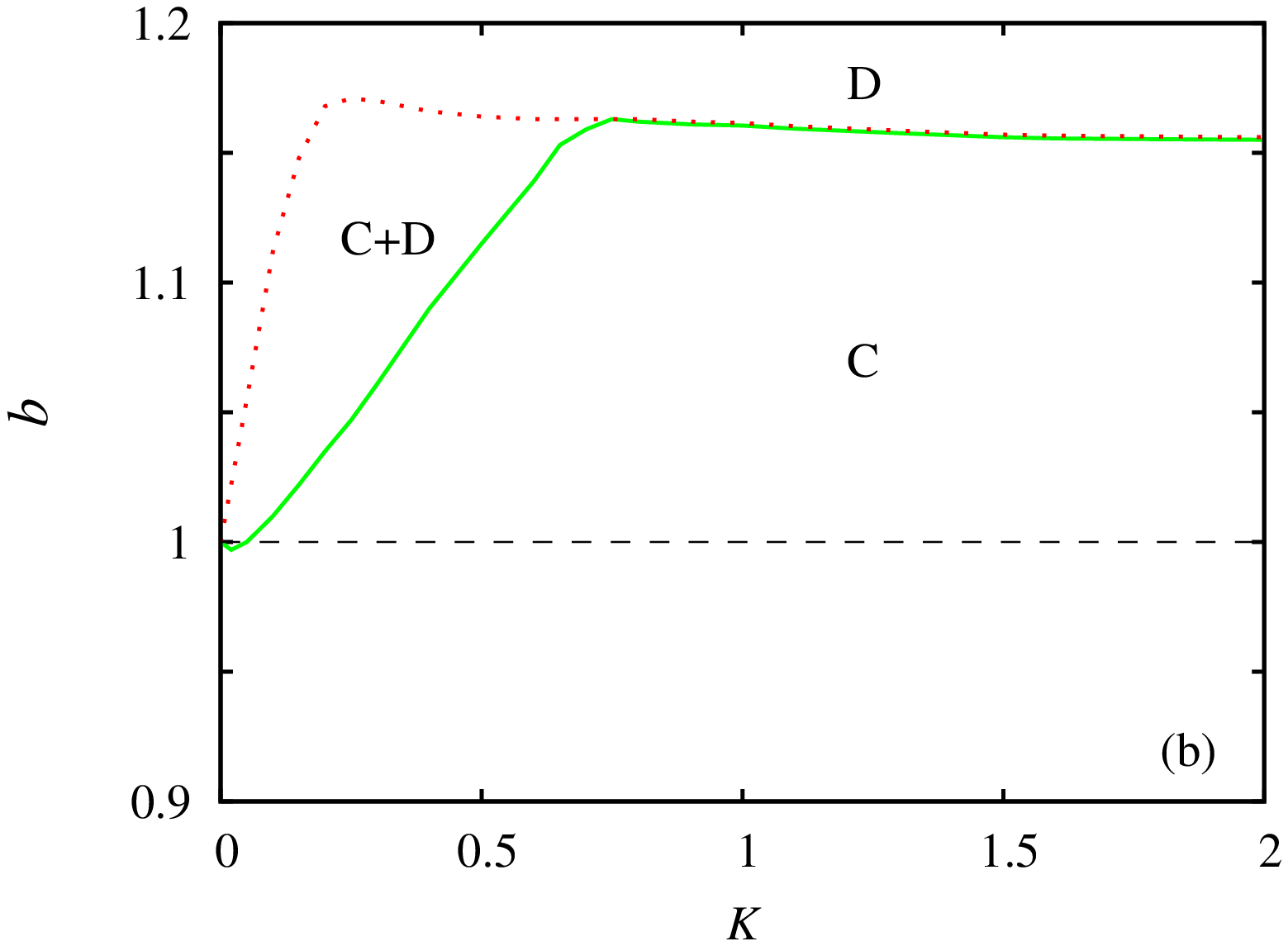,width=7.5cm}}
\caption{(color online) Comparative plots of full $b-K$ phase diagrams for model A obtained by setting H~=~0 [panel (a)] and H~=~10, $w_{min}=0.01$ [panel (b)]. (For easier comparison we used the same $b$ range for both diagrams.) Solid green and dotted red lines mark the borders of all C and all D phases, whereas C+D show the mixed phases where both C and D players are present in the stationary state. Dashes line at $b=1$ shows the limit of weak prisoner's dilemma game.}
\label{phdA}
\end{figure}

\begin{figure*}
\centerline{\epsfig{file=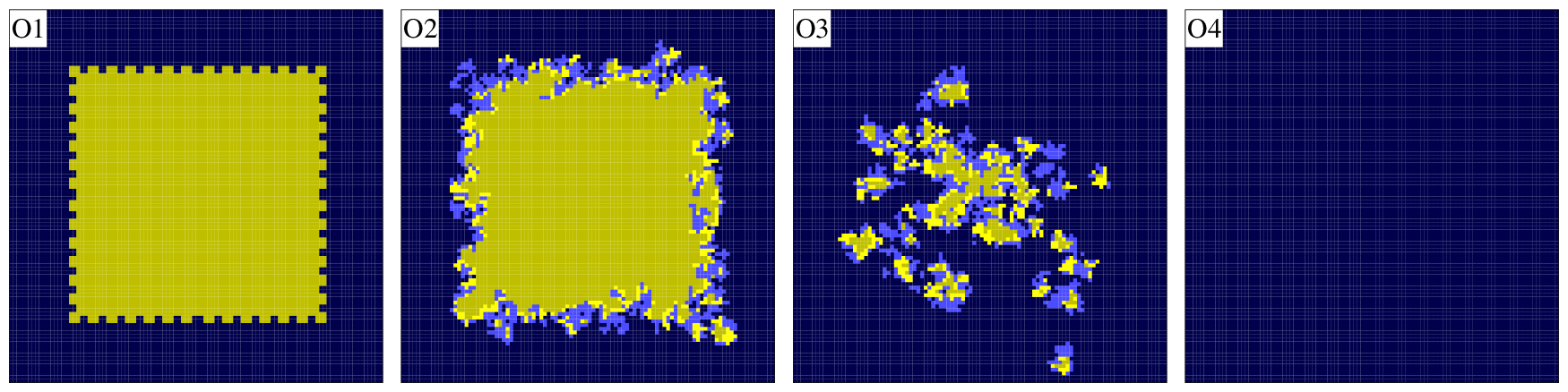,width=17cm}}
\centerline{\epsfig{file=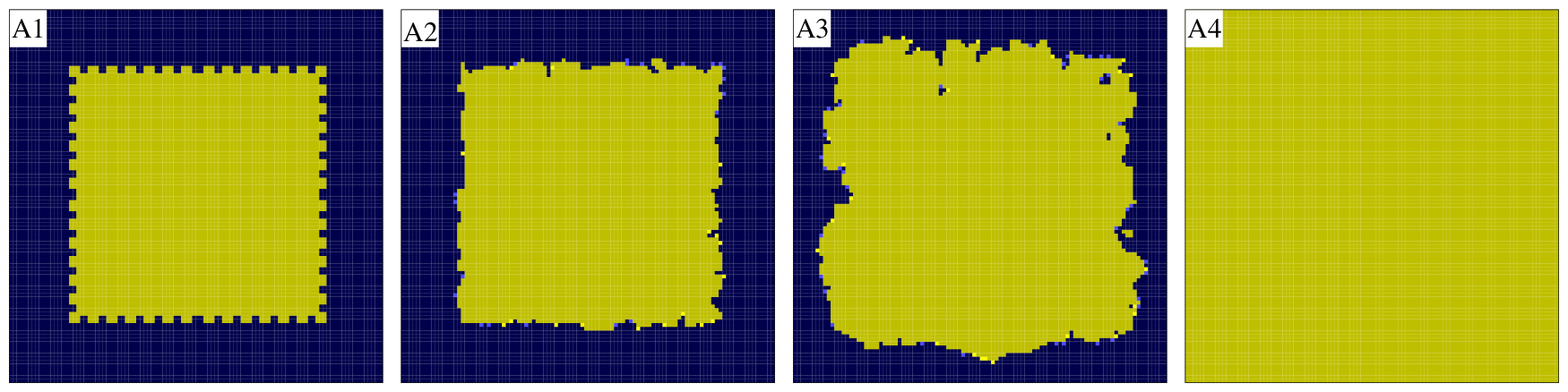,width=17cm}}
\centerline{\epsfig{file=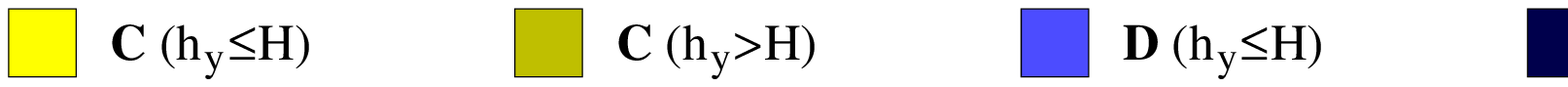,width=17cm}}
\caption{(color online) Time evolutions of a prepared initial configuration on a $100 \times 100$ square lattice for model A at $w_{min}=1$ (upper panel, from O1 to O4) and at $w_{min}=0.01$ (lower panel, from A1 to A4). H~=~10, $b=1.11$, and $K=0.5$ for both panels. As figure legend shows, light yellow (light blue) denotes cooperators (defectors) whose learning capacity is $w_y=w_{max}$ since the elapsed time $h_y$ is shorter than H after strategy adoption. Dark yellow (dark blue) denotes cooperators (defectors) whose learning capacity is $w_y=w_{min}$ as a consequence of large $h_y$ value. In the original model (upper panel) the irregularity of the cooperative border makes the invasion of defectors easy into the bulk of domain. Accordingly, the sliced cooperative domain cannot insure the appropriate environment for network reciprocity to work. In the other case, when $w_{min}=0.01$ (lower panel), the initial irregularity of domain border is smoothed out because cooperators in the bulk can resist against the fast invasion of an intruder defector. Furthermore, cooperators in the bulk can strike back using the support of cooperator neighbors. Snapshots are given after $t=0, 30, 180$, and $450$ iterations for upper panel and after $t=0, 2000, 17000$, and $30000$ iterations for lower panel.}
\label{snapshotsA}
\end{figure*}

In model B the learning capacity of player $y$ changes in time in reverse order, namely 
\begin{equation}
\label{modelB}
w_y = \cases {w_{min},\,\,\,\,\,\,\,\,\,\,\,\,\,\,\,\, \,$if$ \,\,\,h_y \le \,\,$H$ \cr 
              w_{max}=1,\,\,\,\,\, $otherwise$. \cr}
\end{equation}
As we have already noted, the advantage of using step-like functions is to have only one additional parameter for both models, that is the threshold value of elapsed time after strategy adoption when learning capacity changes. More sophisticated functions with smooth time-dependence is also possible, which we will also discuss in the following sections.

The presented Monte Carlo (MC) simulations were obtained on populations comprising $100 \times 100$ to $1600 \times 1600$ individuals, whereby the stationary fractions of strategies were determined within $10^5$ to $10^7$ full MC steps after sufficiently long transients were discarded. It should be noted that the relaxation to stationary state is specially slow for model A. In order to assure suitable accuracy, final results were averaged over up to $5-20$ independent runs for each set of parameter values.

\section{Results for model A}
\label{A}

In this section we present results obtained for model A when the learning capacities of players may reduce after strategy adoption according to Eq.~\ref{modelA}. Figure~\ref{phdA} shows the $b-K$ phase diagrams for H~=~0 [panel (a)] and for H~=~10 [panel (b)]. Note that the case H~=~0 (or $w_{min}=1$) is practically equivalent to the original spatial model where all players have uniform time-independent learning capability \cite{szabo_pre05}. As the comparison demonstrates, the introduction of time-dependent learning capacity results a slight, but unambiguous improvement for cooperation level in dependence of $b$. This impact is more visible for larger H values, but too large H, as we will argue later, would reduce the change from the original model. Interestingly, the phase diagram of model A plotted in Fig.~\ref{phdA}b agrees qualitatively with the $b-K$ diagram when time-dependent teaching activity was supposed \cite{szolnoki_pre09}. Despite this similarity, a completely different pattern formation mechanism can be detected in the present case.

\begin{figure}
\centerline{\epsfig{file=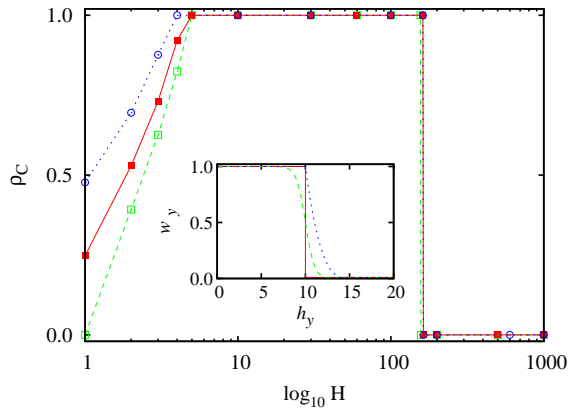,width=7.5cm}}
\caption{(color online) Cooperation level depending on H parameter for model A at different scheduling rules for $b=1.07$ and $K=0.5$. The applied rules are step-function, $\tanh$-, and power-function-like decay as demonstrated by the inset. The corresponding results denoted by closed (red) squares, open (green) squares, and open (blue) circles, respectively.}
\label{time}
\end{figure}

\begin{figure*}
\centerline{\epsfig{file=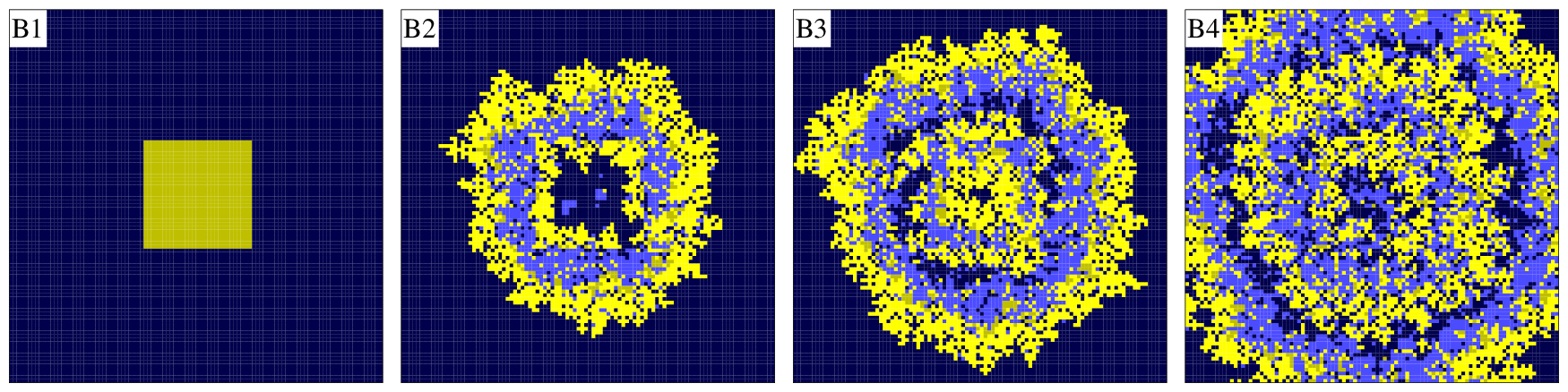,width=17cm}}
\centerline{\epsfig{file=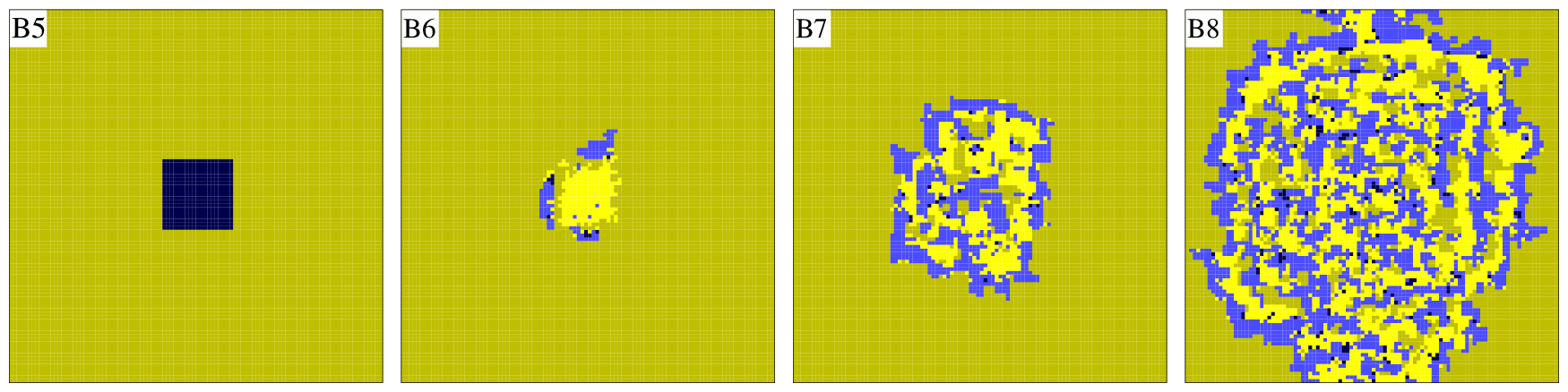,width=17cm}}
\centerline{\epsfig{file=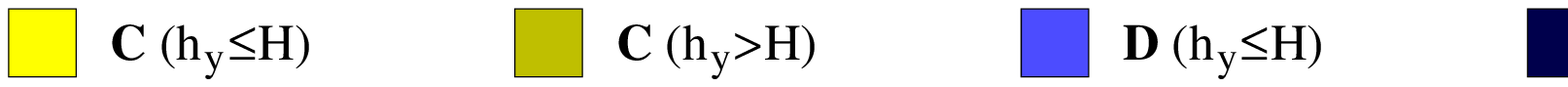,width=17cm}}
\caption{(color online) Time evolutions of prepared initial configurations on a $100 \times 100$ square lattice for model B at $b=1.3$ (upper panel, from B1 to B4) and at $b=0.80$ (lower panel, from B5 to B8). H~=~30, $w_{min}=0.01$, and $K=0.5$ for both panels. As figure legend shows, light yellow (blue) denotes cooperators (defectors) whose learning capacity is $w_y=w_{min}$ since the elapsed time $h_y$ is shorter than H after strategy adoption. Dark yellow (blue) denotes cooperators (defectors) whose learning capacity is $w_y=w_{max}$ as a consequence of large $h_y$ value. The superiority of defectors (cooperators) in upper (lower) panel cannot manifest because a dynamically induced cyclic dominance induces traveling waves that help the suppressed strategy to survive. Snapshots are given after $t=0, 90, 130$, and $210$ iterations for upper panel and after $t=0, 90, 400$, and $880$ iterations for lower panel. Detailed explanation is given in the text.}
\label{snapshotsB}
\end{figure*}

The mechanism that explains the shift of phase borders to larger $b$ values is based on a sort of recovery effect of cooperator domains. The previously mentioned network reciprocity among cooperators can efficiently work if cooperators are aggregated and the border with defector players is smooth. To the contrary, irregular interface, just like well-mixed system, is beneficial to defectors. Consequently, a fault in the interface separating cooperator and defector domains can be a starting point for defectors to initiate invasion into the bulk of cooperator island if $b$ is large enough. This invasion is illustrated on the top panel of Fig.~\ref{snapshotsA} where the sliced cooperative domain cannot resist against the invasion of defectors who can collect larger payoff as a consequence of relatively high $b$ value. Finally, the system terminates into the all D phase.  

In case of time-dependent learning capacity of model A the above described invasion is blocked efficiently in the body of cooperative domain as lower panel of Fig.~\ref{snapshotsA} illustrates. Note that players in the bulk rarely adopt strategy, therefore $h_y$ is large enough to have the minimal $w_{min}$ capacity for them. (Figure~\ref{snapshotsA} demonstrates clearly that players with young strategies, marked by light yellow and light blue, are restricted to the borders separating homogeneous areas.)  Consequently, cooperators with large $h_y$ are reluctant to adopt the intruder defector strategy. On the other side, a player who has just adopted defection at the surface has small $h_y$ hence maximal $w_y=1$ learning capacity. This fact gives a chance for cooperators in the bulk area, who still have supporting cooperator neighbors, to strike back and regain the position of the ``young'' defector. The above described mechanism, as a recovery effect, can smooth the border between defector and cooperator domains, which helps cooperators to survive even at larger $b$ values where cooperators cannot survive in the original model.

The above described effect is more pronounced when H is larger. Too large H, however, has no impact on the final state again. More precisely, if the system is started from a random state and H is large enough ($500\lesssim$~H) then players have no chance to change learning capacity before the stationary state is reached. (In other words, the strategy adoptions happen more frequently than $h_y$ exceeds H.). Accordingly, the evolution to the stationary state is practically similar to the case when there is no time-dependence of learning capacity. This behavior is demonstrated clearly in Fig.~\ref{time}, where the fraction of cooperators is plotted as a function of H parameter. To emphasize the cooperation boosting impact of time-dependent learning activity, we have chosen $b=1.07$ and $K=0.5$ values which would result all $D$ state ($\rho_C=0$) in case of time-independent learning activity (for H~=~0).

Our findings are robust by applying different time profiles for $w_y (h_y)$ function. As the inset of Fig.~\ref{time} shows, we have also applied analytical functions which insure gradual decrease of learning activity. Namely, the applied functions were
\begin{equation}
w_y = \frac{1}{2} [1 - \tanh (h_y - \textrm{H})]
\end{equation}
and
\begin{equation}
w_y = \left( 1 - \frac{h_y - \textrm{H}}{h_{max}} \right) ^{\alpha}
\end{equation}
by keeping the learning activity in the $0.01 \le w_y \le 1$ interval. We used $\alpha=3$ and $h_{max}=5$ to produce data for Fig.~\ref{time}. Naturally, the final state depends on the new parameters but the previously reported basic mechanisms remain intact.

\section{Results for model B}
\label{B}

A surprisingly different behavior can be observed if we reverse the time-dependence of the evolution of learning capacity according to model B, defined by Eq.~\ref{modelB}. As we argued, the temporary block of learning after a strategy adoption can be interpreted in different ways. Thinking about humans, one can easily imagine that a player is satisfied with the recently adopted strategy for a certain period and searches for better strategy just afterwards. More generally, strategy change may involve some cost, hence a player cannot afford to change strategy too frequently.

To reveal the possible consequences, let us follow what happens at the border of $C$ and $D$ domains if cooperators can accidentally invade a site from defectors when $b$ is high. Despite possible high temptation value, the neighboring ``old $D$'' cannot strike back due to low $w_{min}$ value of ``young $C$'' player. Furthermore, the latter actor can resist $D$ until the effective $h_y$ age insures low learning activity. Meanwhile, similar $C \to D$ transition can happen in the neighborhood of the mentioned $C$ player. From this point, $C$ players can support each other and their strategy can spread further in $D$ domains where ''old'' players can adopt the invaders' strategy by having $w_{max}$ learning capacity. In short, ``young $C$'' players can dominate ``old'' defectors.

As time passes, however, $C$ players get older and become vulnerable if $h_y$ exceeds H. Henceforth $D$ can invade ``old $C$'' players again if $b$ is large enough. To sum up, a cyclic ``old $D$'' $\to$ ``young $C$'' $\to$ ``old $C$'' $\to$ ``young $D$'' $\to$ ``old $D$'' dominance evolves. Accordingly, a special pattern formation emerges where the ``fatten in front and slim at back'' mechanism for cooperator domains will produce propagating waves. This is nicely demonstrated in the upper panel of Fig.~\ref{snapshotsB} where cooperators can spread in the sea of defectors despite the fact that $b=1.3$ is much higher than the temptation value where cooperators can survive in the original model with time-independent learning capacity. According to the above mentioned cyclic dominance, the spatial distributions of strategies with specific ``age'' can be easily recognized as shown by the order of colors as ``dark blue'' $\to$ ``light yellow'' $\to$ ``dark yellow'' $\to$  ``light blue'' $\to$ ``dark blue''. It is conspicuous the rare occurrence of ``dark yellow'' (old $C$) players that is a straightforward consequence of high $b$: if a cooperator becomes vulnerable by having $w_{max}$ then it can be promptly invaded by a neighboring defector player.

The same mechanism can be observed if defectors are suppressed due to $b<1$. In this case ``young $D$'' can resist cooperators despite low $b$ value but may invade ``old $C$'' who has $w_{max}$ learning activity. The direct consequence of payoff ranks can only be validated for ``old $D$'' players whose learning capacities are recovered. The lower panel of Fig.~\ref{snapshotsB} illustrates that defectors can survive via propagating waves in the stag-hunt region, too. Naturally, the intensity of dominance between strategies depends on the value of H and the parameters of $b$ and $K$, as well. 

\begin{figure}
\centerline{\epsfig{file=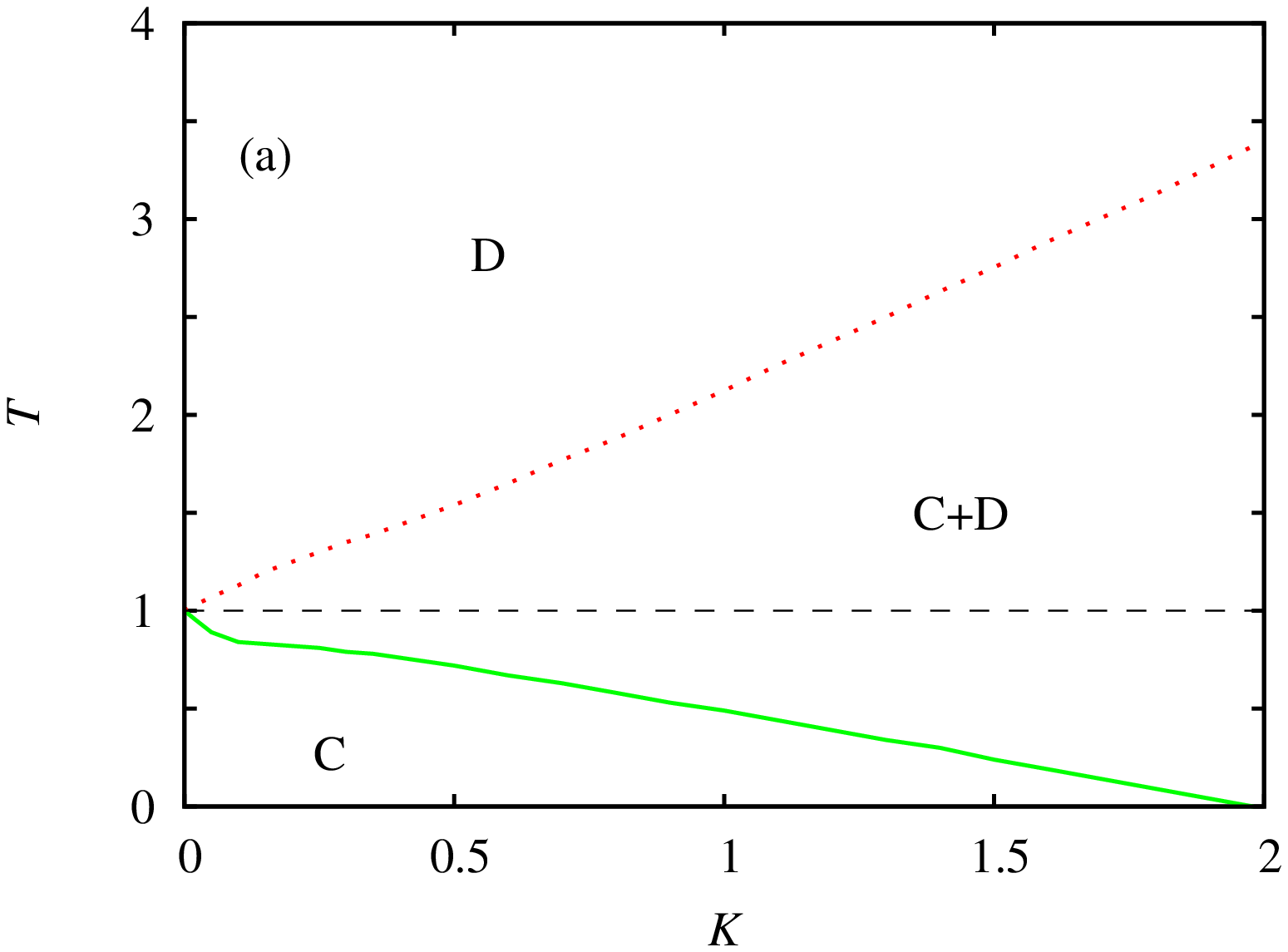,width=7.5cm}}
\centerline{\epsfig{file=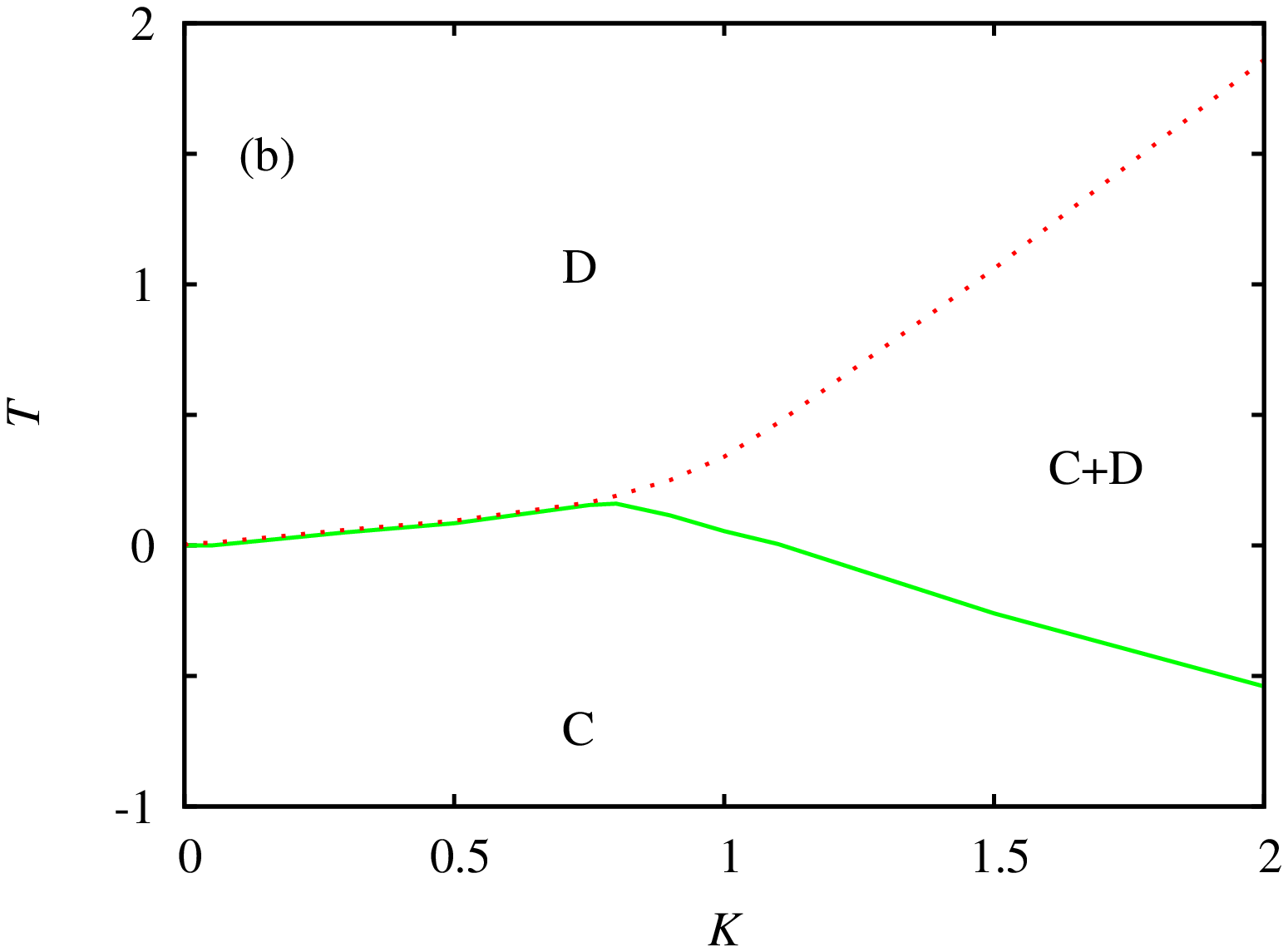,width=7.5cm}}
\caption{(color online) Full $T-K$ phase diagrams for model B obtained for different interaction graph using H~=~10. Panel (a): square lattice at $S=0$. Panel (b): small world graph (with $z=4$) at $S=-1$. In both panels solid (green) and dotted (red) lines mark the borders of all C and all D phases, whereas C+D shows the mixed phase where both strategies are present in the stationary state. Dashes line at $T=1$ shows the limit of weak prisoner's dilemma game.}
\label{phdB}
\end{figure}

As it is known, the general role of $K$ parameter is to moderate the direct consequences of payoff differences originated from the actual rank of payoff elements. Furthermore, as we argued previously, the emerging cyclic dominance is based on a dynamically induced mechanism and depends slightly on payoff elements. Accordingly, the expansion of mixed $C+D$ state, as a result of cyclic dominance, is more vital at larger $K$ values. This expectation is in agreement with the $T-K$ phase diagrams of model B, plotted in Fig.~\ref{phdB}. As the phase diagram indicates, the impact of cyclic dominance is so powerful that mixed phase is extended to such a large $T$ region where the vitality of cooperators cannot be reached by any previously observed cooperator promoting mechanisms. This behavior offers a possible new approach to the basic question why cooperators can survive when payoff elements would predict the opposite solution. Generally, the primary goal is to search for mechanism to strengthen cooperators, hence to extend the survival limit towards cooperator-unfriendly parameter region. On the other hand, the survival of cooperators can be considered as a ``diversity of strategies'' where almost equivalent strategies are maintained via a specific dynamical mechanism.

The coexistence of species due to cyclic dominance is a well-known phenomenon in cyclic Lotka-Volterra models \cite{reichenbach_pre06}, in rock-scissors-paper game \cite{reichenbach_n07}, or in general cyclic competitive system \cite{szabo_pre99}
where at least three \cite{reichenbach_prl08} or more species \cite{szabo_jpa05} are present in the system. In the latter cases the formation of propagating waves is a typical pattern due to cyclic dominance of species \cite{szabo_pre02}. In the present model, there is, however, an important difference from the ``traditional'' cyclic dominated systems. Namely, the relation of states is time-dependent and some states can be transformed to another one without the vicinity of the other state: ``young'' cooperators or defectors will become ``old'' cooperators or defectors as time passes. The time dependence of states, as Glauber-type dynamics, may explain why we cannot observe rotating spirals for model B. 

\begin{figure}
\centerline{\epsfig{file=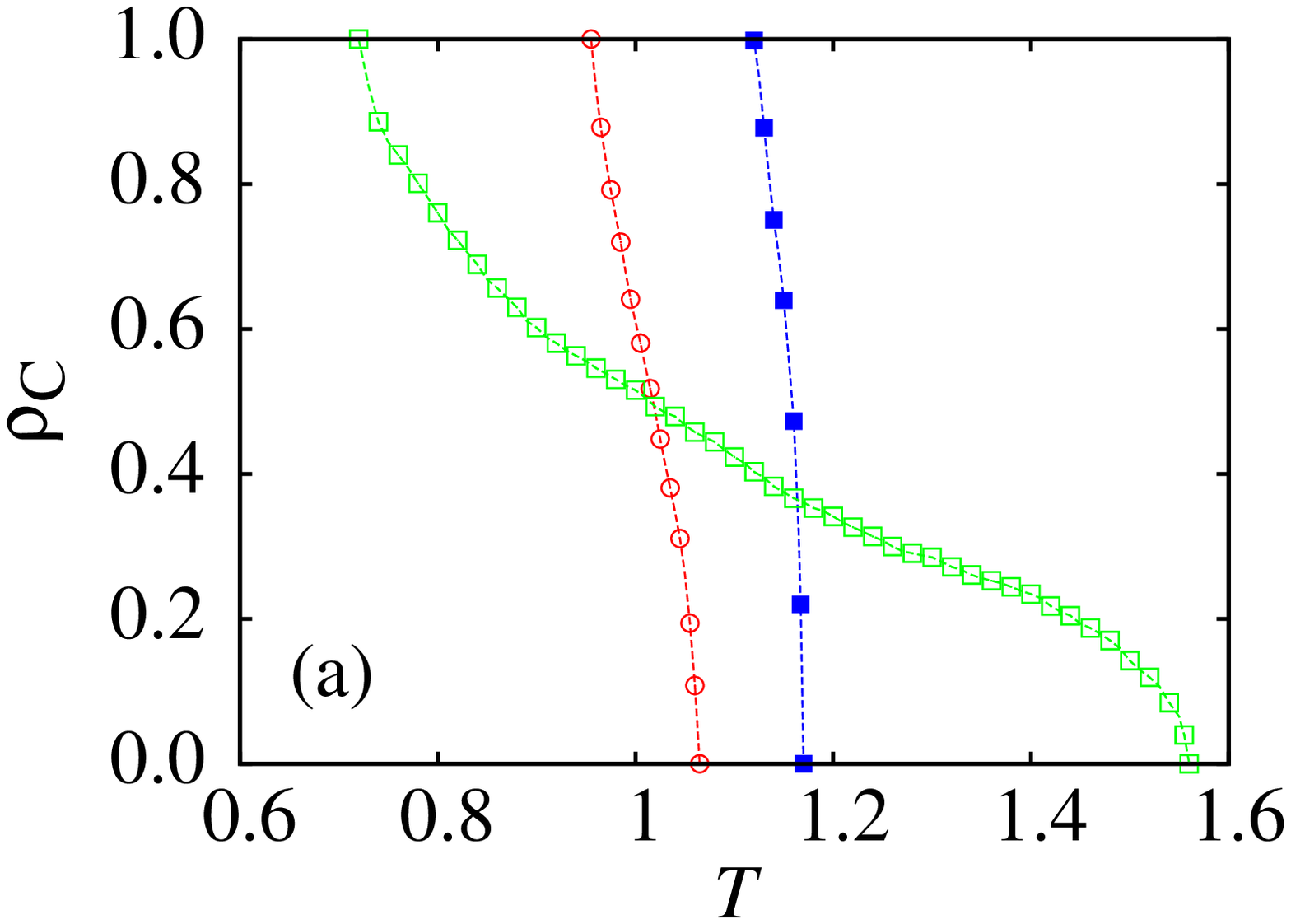,width=4.2cm} \epsfig{file=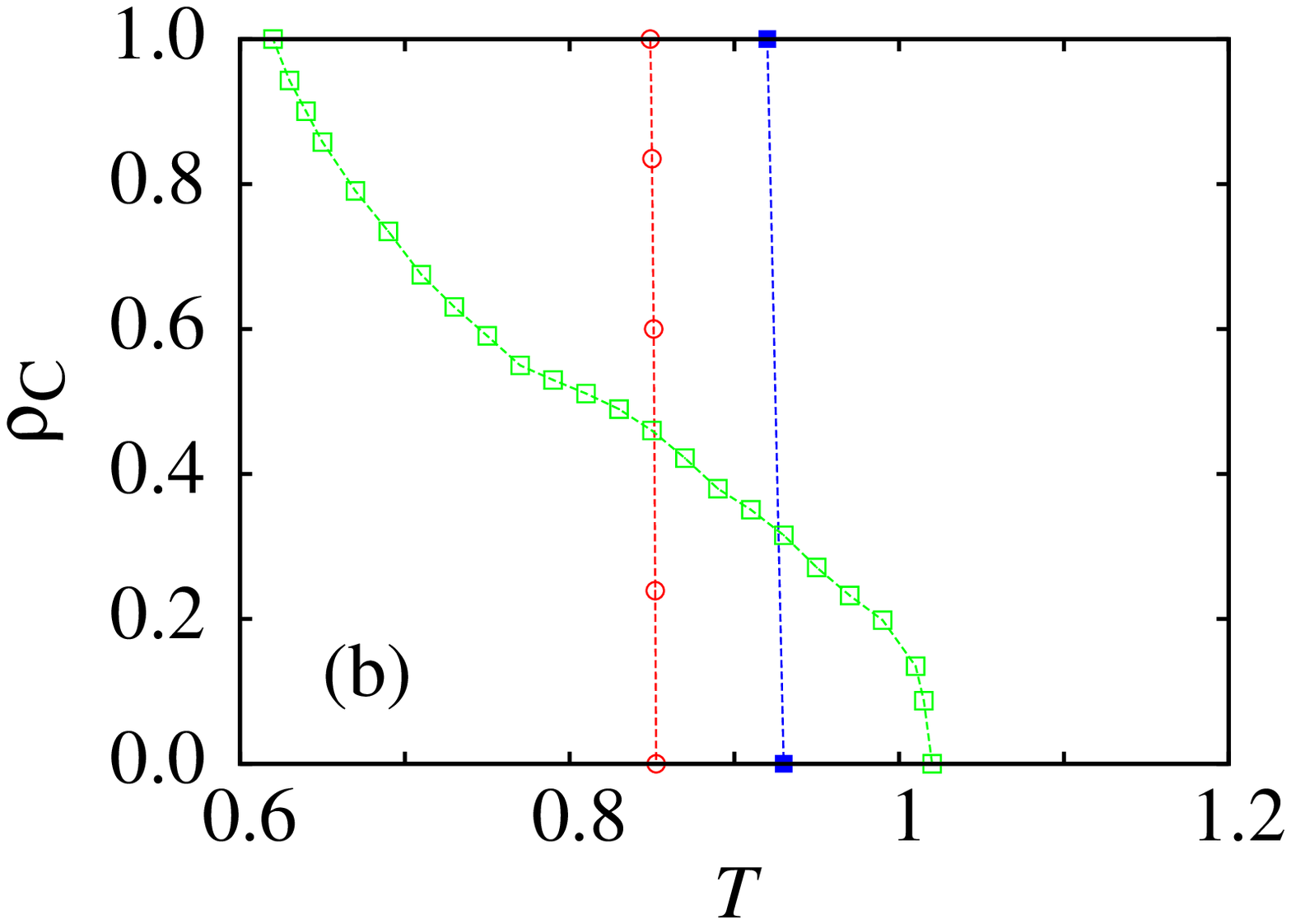,width=4.2cm}}
\centerline{\epsfig{file=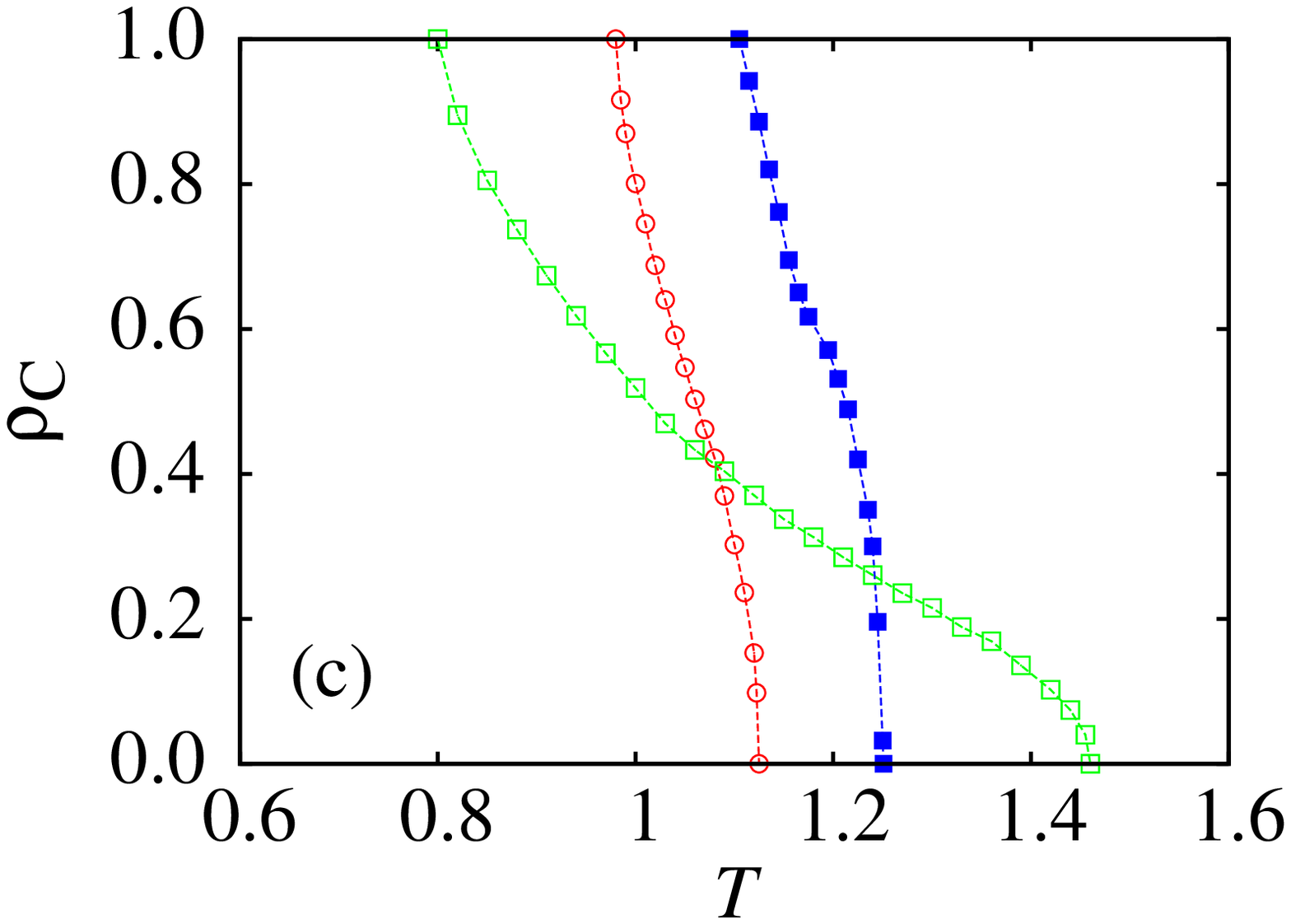,width=4.2cm} \epsfig{file=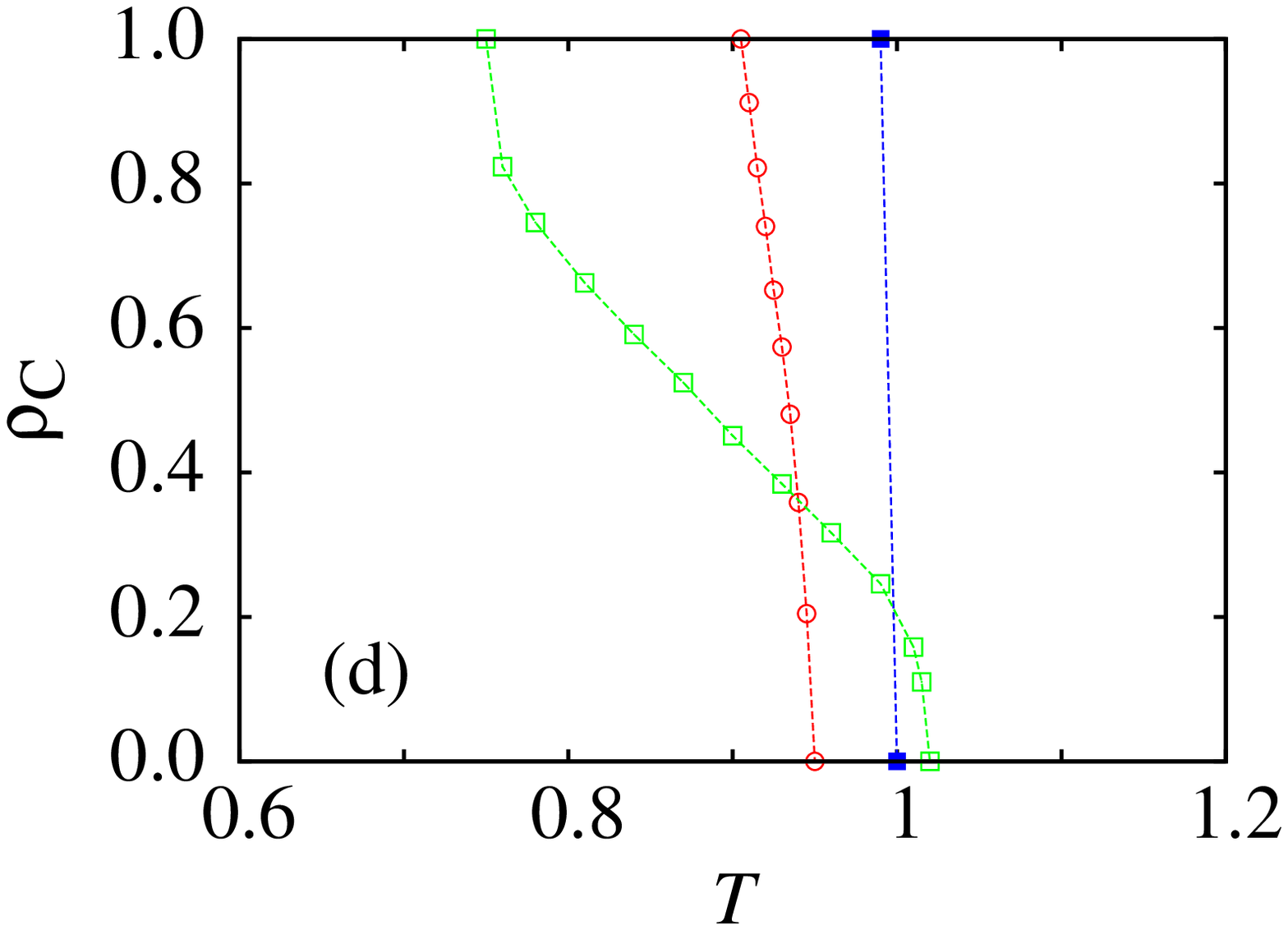,width=4.2cm}}
\caption{(color online) Impact of time-dependent learning activity for different topologies and payoff parametrizations. In all panels open (red) circles depict results obtained with standard version of the game (time-independent teaching activity), while open (blue) and closed (green) squares show stationary $\rho_C$ obtained by model A and model B, respectively. Panel (a): square lattice for $S=0$, (b): square lattice for $S=-0.3$, (c): triangle lattice for $S=0$, and (d): triangle lattice for $S=-0.3$.
In all cases $K=0.5$ and the time schedule parameter is $H=10$.}
\label{all}
\end{figure}

As a further consequence of time-dependent relations of states, there is another type of difference in pattern formation comparing to classical cyclic dominated systems. For rock-scissors-paper game, it was previously demonstrated that the pattern formation depends crucially on the topology of interaction graph of players. More specifically, if lattice topology is replaced by a small-world graph \cite{watts_d_n98} then the time independent strategy density functions become oscillatory as a consequence of short-cuts (long-range interactions) in the connectivity graph \cite{szolnoki_pre04b}. To reveal the possibility of such oscillation, we have also applied random regular graph topology using the same $z=4$ degree of nodes. Despite existing shortcuts the frequencies of strategies were always constant in time in stationary states when the dynamics of model B was applied. As we argued, this deviation may be related to the previously mentioned Glauber-type dynamics in the cyclic dominance. 

Finally, we should stress that our observations are not restricted to specific interaction graph or payoff parametrization of weak prisoner's dilemma game. To demonstrate this, we have plotted sections of phase diagrams in Fig.~\ref{all}, where different lattices and $S$ values were used. As earlier studies suggested \cite{szabo_pre05,szabo_pr07}, interaction graph with substantial clustering coefficient may influence the cooperation level relevantly. To clarify this issue, we have checked triangle lattice and found the same qualitative behavior as for square lattice. Namely, the application of time decaying learning activity improves slightly the cooperation level comparing to the results of time-independent learning activity. In the other case, when the learning activity increases with time after strategy adoption, the mixed C+D phase may expand significantly resulting in a higher chance of cooperators to survive even at large temptation  parameter values.

\section{Summary and discussion}
\label{sum}

In summary, we have studied the impact of time-dependent learning capacities of players within spatial prisoner's dilemma game. There are two significantly different time profiles for learning activity, namely, the skill of a player to adopt alternating strategy may weaken or strengthen as time passes after strategy adoption. By choosing the simplest step-like function, in model A the personal $w_y$ learning capacity changes from $w_{max}=1$ to $w_{min}=0.01$ when the elapsed time exceeds H threshold iterations after strategy adoption. Accordingly in model B, this change happens from $w_{min}$ to $w_{max}$ at a threshold value of H. 

In contrast to the preliminary expectations, these modifications of the standard spatial model with uniform players resulted significant changes in stationary states. The time decreasing learning activity helps cooperator domains to recover the possible intrusion of defectors hence extend the critical temptation to higher values. On the other hand, the temporary restrained learning activity generates a sort of cyclic dominance between defector and cooperator strategies. As a consequence, the spatial pattern is characterized by propagating waves and the resulting mixed $C+D$ state becomes a viable solution even at extremely wide $T$ region starting from stag-hunt to heavy prisoner's dilemma game. It should be also stressed that the microscopic rule is strategy-independent for both models, hence the resulting output is not a straightforward consequence of a direct cooperator supporting strategy update. Our observations are robust and remain valid if we use different interaction graphs or time profiles for strategy learning activity.

Previous studies focused to find cooperator promoting mechanisms that help to extend the limit of parameter space for cooperators to survive. Our work highlights a possible alternative way to keep cooperators alive without postulating additional strategies \cite{hauert_s02,szabo_prl02}. This dynamically induced mechanism, just like migration \cite{helbing_pnas09}, forms new type of patterns in stationary state. Similar dynamically generated (but completely different) effect has already been observed when time-dependent teaching activity (or reputation) of players was supposed \cite{szolnoki_pre09}. The remarkable variety of dynamically induced mechanisms suggests that further pattern formating effects are expected if time-dependent player-specific quantities are supposed in spatial evolutionary game systems.

\begin{acknowledgments}
A. S. acknowledges support from the Hungarian National Research Fund (grant K-73449) and the Bolyai Research Grant.
\end{acknowledgments}

\end{document}